\documentclass[a4paper]{jpconf}
\usepackage{graphicx}
\begin{document}
\title{Many-body correlations in Semiclassical Molecular Dynamics and Skyrme forces
for symmetric Nuclear Matter}

\author{M Papa}

\address{INFN - Sezione di Catania, Via S. Sofia 64, 95123 Catania, Italy}

\ead{massimo.papa@ct.infn.it}

\begin{abstract}
Constraint Molecular dynamics CoMD calculations have been performed for symmetric nuclear matter (NM)
by using a simple effective interactions of the Skyrme type. The set of parameter values reproducing
common accepted saturation properties of nuclear matter have been obtained for different degree of stiffness characterizing the iso-vectorial potential density dependence. A comparison with results obtained in the limit of the Semi-Classical Mean Field approximation performed using the same kind of interaction put in evidence the role played
by the many-body correlations present in the model  explaining also the noticeable differences obtained in the parameter values in the two cases.

\end{abstract}

\section{Introduction}
The description of many-body systems is one of the most difficult problems in nuclear physics  due to the complexity
of this kind of systems which are quantum objects described by a large number of degrees of freedom.
 A large variety of theoretical models have been developed using mean-field based and beyond-mean field  approaches like Density Functional Theory\cite{dft1} and Energy Density Function Theory \cite{ed1,ed2}.
 In these approaches which use the independent particle approximation as a starting point, phenomenological effective interaction like Skyrme and Gogny forces are widely used taking advantage of their simple form\cite{func}. In nuclear structure modeling
  we can quote the Skyrme-Hartree-Fock (SHF) methods, the relativistic mean-field (RMF) approach (as
well as their Bogoliubov extensions), and the Hartree-Fock-Bogoliubov method with the finite-range Gogny force
\cite{stru,rein}.

Complexity still becomes higher when nuclear dynamics, triggered  by nuclear collisions, is studied to understand
the property of nuclear forces far from the stability.
 At the Fermi energy and beyond semiclassical methods become necessary to describe the produced processes in which
 practically all the degrees of freedom are involved. Many-body correlations are responsible for the reorganization of the hot Nuclear Matter(NM) in to clusters. Two main classes of approaches have been developed to handle this complex scenario.
 The first one is based on the Boltzman transport equation including, at the higher energy, also three body collisions term\cite{bnv,vuu,buu}. In this case the theoretical approach is able to describe the time evolution of the one-body distribution function in phase-space. Apart from the collision term,  the Boltzman equation corresponds to the semi-classical limit
 of time-dependent Hartree-Fock equations in phase-space. Different models have been implemented on this ground.
   They essentially differ from each other for the strategy adopted to simulate the collision terms and including the method used to represent the phase-space distribution  starting from the nucleonic degree of freedom (test particle methods).
 As in Harthrre-Fock mean field methods the main ingredient concerning the interaction is the energy density
 which for a Skyrme type phenomenological interaction has a rather manageable form. In particular the local part
 of the interaction produces a simple functional for the potential energy expressed through an algebraic function of the
 density (see also Sec.II). In the early 90  a further development  included
 stochastic forces (typical of Langevin processes) \cite{smf0} in the so called Stochastic Mean Field\cite{smf} model to produce  fluctuations  capable of describing  associated phenomena like mean-field instabilities leading to the cluster production.
 The second line of development of theoretical semiclassical approaches to deal the nuclear many-body problem is represented by the so called Molecular Dynamics models. The point of view of these methods is some sense antithetic
  compared to the ones based on the mean-field concept. However also in these cases, due to their simplicity the same kind of phenomenological
 effective interactions Skyrme and Gogny are widely used. The starting point of all these models is a basic assumption on
 the wave function describing the single nucleon which is represented through a wave packet well localized in
 phase-space with uncertainty satisfying the uncertainty principle.
 The many-body wave fucntion can be expressed through a direct product, leading to the so called Quantum Molecular Dynamics (QMD)-like models \cite{qmd,comd,comdii,qmdcinesi}
 or an anti-symmetrized product like in FMD and AMD \cite{fmd,amd}.
 Variational principles have been used  to obtain the
 equation of motion for the  the many-body functions.
 Within their own approximation schemes these models
  treat the many-body problem without the usage of the mean-field approximation. Many-body correlations
are spontaneously produced  and in general are able to describe the main features concerning the multi-breakup processes \cite{mbreakup} observed in heavy ion collisions at intermediate energy and beyond.
Due to these deep conceptual differences between the mean-field based models and the molecular dynamics ones it
becomes interesting, and necessary in our opinion, to investigate the differences between the energy-density functionals which are produced with these two classes of approaches when one uses, in both the cases, the same kind of effective elementary interaction between nucleons.
The study presented in this work  is performed by using the Constraint Molecular Dynamical Model CoMD \cite{comd,comdii}.
It is a part of a forthcoming more  extended study involving also asymmetric NM.
As it will be shown the presence
of iso-vectorial forces strongly affects the results of this comparison (see also \cite{comdcor}).
The quantitative results obviously will be related to some specificity of the model and to the used effective interaction, but as due to the common features which links most of
the molecular dynamics models (semi classical wave packet dynamics),  the obtained results could assume a more wide meaning at a qualitative level.
The work is organized as  follows: in Sec. II we illustrate the choice of the effective interaction and
 the related total energy functional are introduced.
 In Sec.III we describe the NM simulations. The results are presented in Sec. IV. Sec.V contains the summary and
 the concluding remarks.

\section{The effective interaction and the total energy}
Before illustrating with some detail the model and the NM simulations, in this section we briefly introduce the effective interaction which we use in our calculations. We also present the way in which the related total energy is obtained in the case of the semiclassical mean-field approximation (Se-MFA) and in the case of the semiclassical-wave packet dynamics.

The elementary interaction between two nucleons with spacial coordinates $\mathbf{r}$,$\mathbf{r'}$
and third isospin component $\tau$,$\tau'$  is of the Skyrme type and has the following form:

\begin{equation}\label{1}
V(\mathbf{r},\mathbf{r'})=V^{(2)}\delta(\mathbf{r}-\mathbf{r'}) = \\  \frac{T_{0}}{\rho_{0}}\delta(\mathbf{r}-\mathbf{r'})+\\
\frac{2T_{3}\rho^{\sigma-1}}{(\sigma+1)\rho_{0}^{\sigma}}\delta(\mathbf{r}-\mathbf{r'}) \\
+\frac{T_{4}}{\rho_{0}}F'_{k}(2\delta_{\tau,\tau'}-1)\delta(\mathbf{r}-\mathbf{r'})
\end{equation}
The first term of eq.(1) represents the iso-scalar  contribution to the two-body interaction, the second one
is the usual 3-body effective interaction. For this term the $\rho$ density dependence is modeled through
the parameter $\sigma$. This term appears to be  a generalization of  the Brink-Voutherin \cite{ed2} three-body term corresponding to $\sigma=2$. This generalization is sustained both on the basis of phenomenological
parametrization used in mean-field approaches \cite{vuu} and on more complex microscopic calculations
based on G matrix calculations with realistic interactions.
The third term represents instead the Iso-Vectorial contribution. The form factors $F_{k}$ have the following
expression:
\begin{eqnarray}
F_{k} &=&(\rho/\rho_{0})F'_{k} \\
F'_{1} &=&\frac{2(\rho/\rho_{0})}{1+\rho/\rho_{0}} \\
F'_{2} &=& 1  \\
F'_{3} &=& (\rho/\rho_{0})^{-1/2} \\
F'_{4} &=& (\rho/\rho_{0})^{\gamma-1}
\end{eqnarray}
$\rho_{0}$ is the saturation density value. Also in this case the $F_{k}$ form factors are introduced to reproduce the results of
more complex microscopic calculations based on realistic interaction \cite{pkr,varreal,varreal1,bhf,rmf}
concerning the symmetry energy and reflecting  effects beyond the two-body interaction  . These form
factors have been widely used in Hartree-Fock calculations, in semiclassical BUU calculations and in others
molecular dynamcs models.
In particular, $F_{4}$ will be used, in the limit $\rho/\rho_{0}\sim 1$, to perform calculation with
stiffness parameter $\gamma$ different from the ones related to the others form factors
(in the same limit $F_{1}, F_{2}, F_{3}$ correspond to $\gamma$ values 1.5, 1 and 0.5 respectively).
Finally we observe that for simplicity reason we do not add to this interaction non local effects.
Regarding this assumption we note that some of the
suggested  Skyrme parametrizations have a vanishing or very small terms related to the non locality at the saturation density in the limit of Nuclear Matter (see for example Ref. \cite{skyrmes}).
Moreover as it will be shown  in the following, in CoMD calculations the effective potential which determines the effective force able to govern the motion of the wave packet centroid is already
non-local (sum of Gaussian contributions depending on the intra nucleons distances) being the results of the convolution of the well localized wave packets with $\delta$ function in space.
\vskip 10pt
\noindent
Even if well-known, we shortly recall in the next section the main ingredients of the Se-MFA.
We think this section useful to present in the following
the way in which this assumption is instead broken in the semiclassical wave-packets dynamics.
\subsection{ The Semiclassical Mean Field Approximation}
At a fixed time starting from eq.(1) (we are considering here a stationary problem) , in a rather general way,  we can obtain the  expression for the total energy $W$ related to the potential interaction
by folding the elementary interaction with the two-body density distribution in phase-space
$D(\mathbf{r},\mathbf{r'},\mathbf{p},\mathbf{p'})$. By taking into account the usual truncation condition on $D$ typical of the mean-field approximation, we can write:
\begin{eqnarray}
D&=&D_{1}(\mathbf{r},\mathbf{p})\cdot D_{1}(\mathbf{r'},\mathbf{p'})
\end{eqnarray}
where $D_{1}$ is the one-body distribution. Therefore by using eq.(1) for $W$ we get:
\begin{equation}\label{XX}
W =\frac{1}{2}\int V(\mathbf{r},\mathbf{r'})D(\mathbf{r},\mathbf{r'},\mathbf{p},\mathbf{p'})d\mathbf{r}d\mathbf{r'}d\mathbf{p}
d\mathbf{p'}\\
=\frac{1}{2}\int V^{(2)}D_{1}(\mathbf{r},\mathbf{p})D_{1}(\mathbf{r},\mathbf{p'})
d\mathbf{r}d\mathbf{p}d\mathbf{p'}
\end{equation}
Taking in to account that by definition $\int D_{1}(\mathbf{r},\mathbf{p)d\mathbf{p}}=\rho(\mathbf{r})$ we obtain:
\begin{eqnarray}
W &=&\frac{1}{2}\int V^{(2)}\rho^{2}d\mathbf{r}
\end{eqnarray}
$\frac{1}{2} V^{(2)}\rho^{2}$ appear to be the energy density associated to the potential energy from which we
can obtain the related binding energy per nucleon due to the potential interaction $E_{pot}$
\begin{equation}\label{XX}
E_{pot}=U_{twb}+U_{trb}+U_{asy}\\
=\frac{1}{2}V^{(2)}\rho=\frac{1}{2}\frac{T_{0}\rho }{\rho_{0}}
+\frac{ T_{3}\rho^{\sigma} }{ (\sigma+1)\rho_{0}^{\sigma} }
+\frac{1}{2}T_{4}F_{k}(\rho)\beta^{2}
\end{equation}
$\beta=\frac{\rho_{n}-\rho_{p}}{\rho}$ represents the charge/mass asymmetry parameter evaluated from the neutron and proton density
$\rho_{n},\rho_{p}$ respectively.
The total binding energy $E$ is obtained by adding the kinetic contribution coming from the Fermi motion
\begin{eqnarray}
 E &=& E_{pot}+E_{kin}^{F} \\
E_{kin}^{F} &=&\frac{3}{5}\frac{\hbar^{2}}{2m_{0}}(\frac{3\pi^{2}\rho}{2})^{2/3}[1+\frac{5}{9}\beta^{2}]
\end{eqnarray}
In $E_{kin}^{F}$ $\beta$ terms of order greater than two are neglected.
In particular we see how the iso-vectorial vectorial forces with strength proportional to $T_{4}$
contribute  to the symmetry energy $E_{sym}$ depending on the asymmetry parameter $\beta$. For quadratic form in
$\beta$ we get:
\begin{eqnarray}
E_{sym}(\rho)&=&e_{sym}\beta^{2}
\end{eqnarray}
\begin{equation}\label{XX}
e_{sym} = \frac{1}{2}(\frac{\partial^{2}E}{\partial\beta^{2}})\\ =\frac{1}{6}\frac{\hbar^{2}}{m_{0}}(\frac{3\pi^{2}\rho}{2})^{2/3}
+\frac{1}{2}T_{4}F(\rho)
\end{equation}
The common accepted bulk value of $e_{sym}$ at the saturation density is about 30 MeV even if relativistic Hartree-Fock models can predict
higher values up to about 40 MeV \cite{barrep,baorep}. However, still under study is the density dependence of this quantity which is able to affect neutron-skin  thickness in nuclei and the Giant Monopole Resonances  \cite{barrep,baorep,tsang}.
Finally we note that the  structure of the obtained energy-density functional in
Se-MFA  makes independent the choice
of the parameter describing the main properties of the symmetry energy  from the other ones  which instead are fixed from the
saturation properties of the symmetric nuclear matter.
We will show in the next section that this is no longer true in the CoMD approach.

In molecular dynamics approaches a strong assumption is done on the wave functions describing the nucleonic
degree of freedom. It is  commonly assumed that the wave function is a gaussian wave packet with fixed width $\sigma_{r}$ in coordinate space.
The centroid in phase-space is indicated with $\mathbf{r}_{i}$, $\mathbf{p}_{i}$
\begin{equation}
\phi_{i}=\frac{1}{(2\pi\sigma_{r}^{2})^{\frac{3}{4}}}exp[-\frac{(\mathbf{r}-\mathbf{r}_{i})^{2}}{2\sigma_{r}^{2}}
-i\frac{\mathbf{r}\mathbf{p}_{i}}{\hbar}]
\end{equation}
The Wigner transform of $\phi_{i}$ is
\begin{equation}
f_{i} = \frac{1}{(2\pi\sigma_{r}\sigma_{p})^{3}}exp[-\frac{(\mathbf{r}-\mathbf{r}_{i})^{2}}{2\sigma_{r}^{2}}
 -\frac{(\mathbf{p}-\mathbf{p}_{i})^{2}}{2\sigma_{p}^{2}}]
\end{equation}
the widths in momentum and space satisfy the minimum uncertainty principle condition $\sigma_{r}\sigma_{p}=\frac{1}{2}\hbar$.
Another assumption concerns the N-body Wigner distribution that in CoMD model \cite{comd}(like in QMD approach\cite{qmd}) is a direct product of the single particle distributions. In this case therefore for a system
 formed by A nucleons the one-body and 2-body distributions above introduced have a multi-component structure that is:
\begin{eqnarray}
 D_{1}(\mathbf{r},\mathbf{p})&=&\sum_{1}^{A}f_{i}(\mathbf{r},\mathbf{p})\\
 D(\mathbf{r},\mathbf{p},\mathbf{r'},\mathbf{p'})&=&\sum_{i\neq j=1}^{A}f_{i}(\mathbf{r},\mathbf{p})f_{j}(\mathbf{r'},\mathbf{p'})
\end{eqnarray}
From the above relations we can see that in general $D\neq D_{1}(\mathbf{r},\mathbf{p})D_{1}(\mathbf{r'},\mathbf{p'})$.
In particular, for the case in which these distributions are expressed as a sum of different localized components inside a volume $V_{g}$, we can indicate with $a_{g}$  the number of components which give non negligible
 contributions in $V_{g}$. The relative difference associated to eq.(18) $1-\frac{D_{1}(\mathbf{r},\mathbf{p})D_{1}(\mathbf{r'},\mathbf{p'})}{D}$ can be estimated to be of the order of $1/a_{g}$ which corresponds to the ratio between diagonal ($i=j$) and off-diagonal elements ($i\neq j)$
  within the ensemble of $a_{g}(a_{g}-1)$ terms. In semiclassical mean-field models
$a_{g}$ can be enough high to make the difference negligible, in fact the single particle distribution usually spreads over the whole system (test particles methods) and therefore the truncation condition eq.(18) can be  retained valid \cite{bnv,vuu,buu}. On the contrary  this is not surely the case for the molecular dynamics approaches  for which the typical  spreading volume is of the order of 2-10 $fm^{3}$.
Localization and therefore coherence of  the wave-packed used
to describe the single-particle wave-functions allows to keep memory of the two-body nature
of the inter-particles interaction, and at same time, allows for the spontaneous
appearance of the clustering phenomena in simulations concerning low-density and excited portion of nuclear matter.
With these assumptions on the 2-body phase-space distribution, taking in to account the properties of the $\delta$ function, we can obtain the explicity expression for the
different terms concerning the total energy; for the two-body isoscalar contribution $W_{twb}$ we get:
\begin{eqnarray}
W_{twb}&=&\frac{T_{0}} {2\rho_{0}(4\pi\sigma_{r}^{2})^{3/2)} }\sum_{i\ne j=1}^{A}
exp[-\frac{ (\mathbf{r}_{i}-\mathbf{r}_{j})^{2} }{4\sigma_{r}^{2}}]\\
W_{twb}&=&\frac{T_{0}}{2\rho_{0}}\sum_{i=1}^{A}S_{v}^{i}\\
S_{v}^{i}&=& \sum_{j\neq i=1}^{A}\frac{1}{(4\pi\sigma_{r}^{2})^{3/2}}exp[-\frac{(\mathbf{r}_{i}-\mathbf{r}_{j})^{2}}{4\sigma_{r}^{2}}]
\end{eqnarray}
In the above expression $S_{v}^{i}$ is the normalized sum of the Gaussian terms and it represents just a measure of the overlap
between the nucleonic wave-packets. Its two body character is quite explicit.
In the calculations concerning the NM simulation that we will illustrate in the next sections , the large
number of particles $A$ involved in the system allows us to write the above quantity in a simpler way by
introdcing the average overlap per nucleon $\overline{S_{v}^{i}}=\overline{S_{v}}$ whose dependence on the particle index in the ideal case can be omitted:
\begin{eqnarray}
W_{twb}&=&\frac{T_{0}A}{2\rho_{0}} \overline{S_{v}} \\
E_{twb}&=&\frac{W_{twb}}{A}=\frac{T_{0}}{2\rho_{0}}\overline{S_{v}}
 \end{eqnarray}
By comparing the expression obtained for $E_{twb}$ in the two different appraches (eq.(10) and eq.(23))
 we note a formal analogy where the variable $\rho$ is substituted by the overlapp integral $\overline{S_{v}}$ per nucleon.
We however observe that this analogy is only formal, this aspect will be discussed in some detail in the next
subsection.
Concerning the three-body term, according to the evaluations reported in reference \cite{qmd} and taking in to
account the previous obervations we get:
\begin{eqnarray}
E_{trb}&=&\frac{T_{3}}{(\sigma+1)\rho_{0}^{\sigma}}\overline{S_{v}}^{\sigma}
\end{eqnarray}
For the term related to the iso-vectorial interaction and for the most simple case $F'=1$ in the limit $A,N,Z >>1$ ($N$ and $Z$ represent the number of neutron and protons) we obtain:
\begin{eqnarray}
W^{isv}&=&\frac{T_{4}}{2 \rho_{0}}(N^{2}\tilde{\rho}^{nn}+Z^{2}\tilde{\rho}^{pp}-2NZ\tilde{\rho}^{np})\\
\tilde{\rho}^{nn}&=&\frac{1}{(4\pi\sigma_{r}^{2})^{3/2}N^{2}}\sum_{i\neq j\in\textsf{N}}
exp[-\frac{ (\mathbf{r}_{i}-\mathbf{r}_{j})^{2}}{4\sigma_{r}^{2}}]\\
\tilde{\rho}^{np}&=&\frac{1}{ (4\pi\sigma_{r}^{2})^{3/2}Z^{2} }\sum_{i\neq j\in\textsf{Z}}
exp[-\frac{ (\mathbf{r}_{i}-\mathbf{r}_{j})^{2}}{4\sigma_{r}^{2}}]\\
\tilde{\rho}^{np}&=&\frac{1}{(4\pi\sigma_{r}^{2})^{3/2}2NZ}\sum_{i\neq j\in\textsf{NZ}}
exp[-\frac{ (\mathbf{r}_{i}-\mathbf{r}_{j})^{2}}{4\sigma_{r}^{2}}]
\end{eqnarray}
where $\tilde{\rho}^{cc'}$ with $cc'$ equal to $nn$, $pp$ and $np$ represents the overlap integral per
couples of neutrons, protons and neutron-proton.
A more convenient form for the above expression is obtained by introducing the two followng quantities:
\begin{eqnarray}
\tilde{\rho}&=&\frac{N^{2}\widetilde{\rho^{nn}}+Z^{2}\widetilde{\rho^{pp}}}{N^{2}+Z^{2}}\\
\alpha&=&\frac{\widetilde{\rho^{np}}-\widetilde{\rho}}{\widetilde{\rho}}
\end{eqnarray}
for symmetric NM $N=Z$ and we get:
\begin{eqnarray}
W_{isv}^{C}&=&-\frac{T_{4}}{4\rho_{0}}A^{2}\tilde{\rho}\alpha\\
E_{isv}^{C}&=&-\frac{T_{4}}{4\rho_{0}}F'_{k}(\overline{S_{v}})\tilde{\rho}_{A}\alpha=E_{bias}
\end{eqnarray}
where $\tilde{\rho}_{A}\equiv A\tilde{\rho}$.
 The expression in eq.(32) also contains a generalization to the cases
 in which we  use the  generic form factors $F'_{k}$. Here $F'_{k}$ keep the same  functional form in  $\overline{S_{v}}$ using the formal analogy $\frac{\rho}{\rho_{0}}\rightarrow \frac{\overline{S_{v}} }{\overline{S_{v,0}}}$ above discussed.
From eq.(32) we obtain that for $\alpha \neq 0$ (as CoMD calculations predicts, see next sections) the iso-vectorial force produces a term  which we
 name $E_{bias}$. The related  effect is not-negligible if compared to the balance of the different terms appearing in the expression of the total energy.
The correlation coefficient $\alpha$ by definition (see eq.(30)) represents the difference in percentage
of the overlap between the neutron-proton couples  from the one related to the couples formed by homonym nucleons.
It also depends on the strength of the iso-vectorial forces ($T_{4}$ parameter).

Finally, the kinetic contribution $E_{kin}$ is obtained in a self-consistent way for ground state configurations
 by applying the cooling-warming procedure coupled with the constraint on the occupation numbers (see Ref.\cite{comdii}) given by the Pauli principle.
 The total energy per nucleon is therefore obtained by adding all the contributions
 discussed in this section and it will be indicated  as $E^{C}=E_{pot}(\overline{S_{v}}$, $\alpha$, $\tilde{\rho}_{A})+E_{kin}(\rho)$. In particular we note that, at this level, it
  depends on the new defined primary quantity $\overline{S_{v}}$, $\alpha$, $\tilde{\rho}_{A}$.
In the following section we will try to relate these quantities to more fundamental ones that are
the density $\rho$ and the spatial correlation function $\nu$ between nucleon pairs.

\subsection{Overlap integrals and spacial correlations}
In a rather general way and as suggested from CoMD calculations (see Fig.1),
for an uniform many-body system at a fixed density we can introduce the probability $p$ to have a particle in the volume $dV_{1}$  localized in $\mathbf{r}_{1}$ and a second one in the volume $dV_{2}$ localized in $\mathbf{r}_{2}$.
Due to the uniformity condition $p$ depends only on  $r=|\mathbf{r}_{1}-\mathbf{r}_{2}|$.
$p$ can be expressed in the following form:
\begin{eqnarray}
p&=&1\pm k_{0}\nu(r)
\end{eqnarray}
with $\nu(0)=1$ and  $k_{0}\geq 0$.
Moreover lim$_{r\rightarrow \infty}\nu(r)=0$, that is: no spacial correlations can be expected for very distant particles.
Non zero values of $\nu$ can be instead expected at relatively small distances due both to the interaction (for an attractive interaction the positive sign must be considered in eq.(33)) and to the symmetry of the many-body wave function for quantal systems of identical particles (in the case of identical Fermions we must consider the sign minus and $k_{0}=1$ ).
For a classical system of non interacting  particles we have a vanishing correlation effect $p=1$ ($k_{0}=0$).
For our aims, we need to evaluate a normalized probability $P=cp$ in such a way:
 \begin{eqnarray}
 \int_{V}PdV&=&4\pi c \int_{0}^{r'}p(r)r^{2}dr=A\rightarrow c=\frac{A}{V-Vc}
\end{eqnarray}
$Vc$ represents the volume in which the well localized correlation function $\nu$ is different from zero.
This volume is always finite and of the order of the ${\sigma_{r}}^{3}$ in our model calculations. So that
in the limit $V\rightarrow \infty$ we get $c=\rho$ and :
 \begin{eqnarray}
 P(r)&=&\rho(1\pm k_{0}\nu(r))
\end{eqnarray}
In Fig.1, just as an example, we display results of calculations for a given set of parameters for the Skyrme interaction (see the next section). The calculations show the probability to find two nucleons at a distance $r$ in case of Pauli blocked couples $P_{1}$ (red points, neutron and proton couples with same spins), for the case of un-blocked proton and neutron couples  $P_{0}$ (black point) and finally
for  neutron-proton couples $P_{np}$ (blue points).
 \begin{figure}
\includegraphics[scale=0.6]{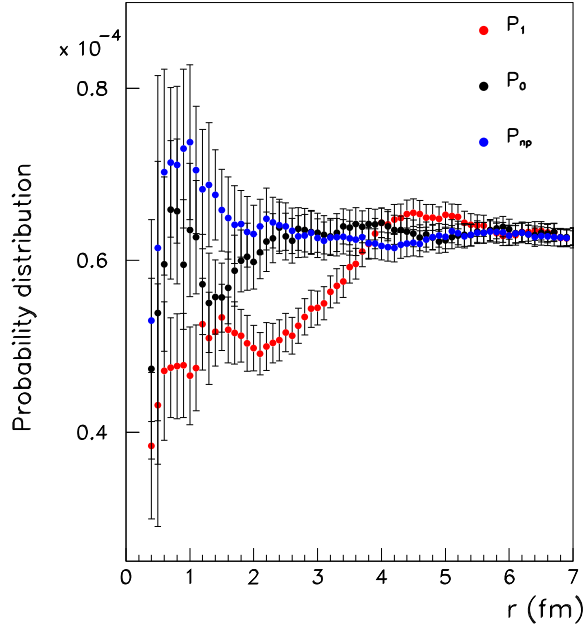}
\centering
\caption{\label{fig2} Probability distribution $P_{1}$ to find two identical nucleons at a relative
distance $r$. In the same figure $P_{0}$ represents the probability evaluated for neutron and proton couples
with opposite spin. Finally $P_{np}$ represents the probability obtained for neutron-proton couples. (Color)}
\end{figure}

The calculations are referred to  a symmetric portion
of nuclear matter consisting of a sphere containing 2000 nucleons at a density of about 0.17$fm^{-3}$. The calculations  include also the iso-vectorial potential energy  ($T_{4}=59$ MeV in this example ). The  probability distribution has been estimated by taking in to account only couples of nucleons
in which at least one of the two nucleons is localized inside a smaller sphere, centered at the origin,
with a radius of 4 fm. In this way we can
neglect  surface effects on this quantity. The error bars indicates the uncertainty related to the statistics of the simulations.
We remark that in all the calculations concerning the present work in the range of density explored
no cluster formation  has been detected by our numerical procedure which was
applied at each time step to check  for the existence of aggregation processes. Only in the lower limit $\rho=0.7\rho_{0}$ and for only
part of the time, the formation of a big cluster with a mass about 98 \% the total one and light particles have been observed.
From the figure it is clearly seen the depletion of the $P_{1}$ distribution at small distances due to Pauli constraining in CoMD. For the other distributions an enhancement is produced  due to the attractive effect of the two-body interaction. In particular we note that this enhancement is more pronounced for $P_{np}$ as compared to
$P_{0}$; this is clearly an effect due to the iso-vectorial forces which generate a stronger attraction for neutron-proton couples.
Now from eq.(35)  we can obtain an expression for $\overline{S_{v}}$ as:
\begin{equation}\label{xx}
 \overline{S_{v}}=\frac{1}{(4\pi\sigma_{r}^{2})^{3/2}}\int P(r)exp[-\frac{r^{2}}{4\sigma_{r}^{2}}]dV\\
 =\rho(1\pm I)
\end{equation}
 \begin{eqnarray}
 I&=&\frac{1}{\sqrt{4\pi}\sigma_{r}^{3}}\int_{0}^{\infty}r^{2}k_{0}exp[-\frac{r^{2}}{4\sigma_{r}^{2}}]\nu(r)dr
\end{eqnarray}
From the above equations we see that even if the overlap integral per nucleon is closely related to the density
$\rho$, it admits correction terms related to the spacial correlation through the integral $I$.
 In general, for function $\nu(r)$ localized within a distance of the order of $l$ the value of $I$  decreases with the ratio $x=\frac{\sigma_{r}}{l}$: for example, for an exponential profile $e^{-\frac{r}{l}}$
the correction terms go to zero like $e^{-\frac{ \sigma_{r}^{2}} { 2l^{2}}}$.
In other words, according to what observed in the previous section,  when the single particles space distribution
is enough large an averaging of the spacial correlations effect is obtained and the CoMD  functional tends to the one represented in eq.(10) which is typical of the semi-classical mean-field approximation in transport theory.
On the contrary for well localized  wave-packets with $\sigma_{r}$ of the order of 1-2 $fm$, $I$ is different from zero and reflects the behavior of $\nu(r)$ at small distances.
The above relations can be also generalized for a multi-component system characterized by a charge/mass parameter different from zero.

 The determination of the correlation functions $\nu(r)$ and of the related integral $I$ is a problem which can not be solved in a general way. Some special cases are well known. They concern  the study of the density fluctuations in the hydrodynamical limit valid for large distances compared to the mean free-path  in non-equilibrium cases
(see also \cite{smf0,smf}).However in this limit simultaneous spacial correlations are supposed to be zero at small distances.
At short distances, comparable with the range of the effective interaction,
as in our case,  approximations schemes can be developed  actually corresponding to a power decomposition
in $\rho$ as we will perform in the next section.

\section{ The nuclear matter simulation}
As mentioned in the introduction the main goal of the present work is
to understand also at a quantitative level and for a given simple form of the effective interaction,
 what are the consequences produced by the correlations above discussed  by taking as a reference point
the results obtainable in the framework of the Se-MFA approach.
The study will be performed in a narrow region of density around the saturation density $\rho_{0}$.
To this aim in the following we will try  to find the set of parameters for the effective interaction
which reproduce in both the cases some of the commonly accepted saturation properties of symmetric nuclear matter.
In particular we will refer to  $\rho_{0}$=0.165 $fm^{-3}$, binding energy $E(\rho_{0})=-16$ MeV, compressibility modulus for symmetric nuclear mater $K_{NM}(\rho_{0})=220$ MeV.
The strength of the iso-vectorial interaction $T_{4}$ will be chosen in such way to produce
 a symmetry energy value $e_{sym}(\rho_{0})\simeq 28.6$ MeV in the case of the Se-MFA, i.e.
 $T_{4}=32$ MeV.
\noindent
 From the functional $E$ reported in eqs.(10-12)
we can find easily the parameter values
satisfying the above conditions by solving the system formed by the following  set of equations for symmetric NM:
\begin{eqnarray}
\rho_{0}&=& 0.165 \\
 E(\rho_{0})& =& -16 MeV\\
 \frac{d E}{ d \rho }/_{\rho_{0}}&=&0\\
 9\rho_{0}^{2}\frac{d^{2} E}{d^{2} \rho}/_{\rho_{0}}&=&K_{NM}(\rho_{0})=220 MeV
 \end{eqnarray}
In concrete cases due to the finite steps with which we perform the variation on the parameters the
 values of $E(\rho_{0})$ and $K_{NM}(\rho_{0})$ are obtained within $\pm 0.5\%$.
Within the above specified uncertainty the solution gives the following values for the parameters:
$T_{0}=-263\pm 1.3$ MeV, $T_{3}=208\pm 1$ MeV and $\sigma=1.25\pm 0.02$.

\subsection{NM calculations and CoMD model}
The evaluation of the total energy per nucleon $E^{C}$  related to the CoMD calculations requires
the solution of the many-body problem using the equation of motion regulating the wave-packet dynamics.
Details on the numerical procedure for the CoMD calculations
are described in ref.\cite{comd,comdii}.

At the different densities changing between 0.7-1.2$\rho_{0}$ with  steps equal to $0.1\rho_{0}$, the calculations have been performed by
enclosing a relatively large number of particles $A_{1}=1600$ and $A_{2}=3560$ in  spherical volumes of radii $R=r_{0}A^{1/3}$ with $r_{0}=(\frac{3}{4\pi\rho_{0}})^{1/3}$. Particles trying to escape from the spheres
have been re-scattered inside through an elastic reflection at the surface.
For symmetrical configuration, starting from the parameter values minimizing the functional $E$ in eq.(10-12) we have searched for the stationary minimum energy conditions by applying the cooling-warming procedure coupled with the constraint related to the Pauli principle as described  in Ref.\cite{comd}.
Calculations  have been performed for the two systems
having  number of particles equal to $A_{1}$ and $A_{2}$.
The value of $T_{4}$ has been fixed to 32 MeV and
the calculations have been performed for different values of the stiffness parameter $\gamma$.
From the minimum energy configurations  we have
evaluated  the related intensive quantities $\overline{S_{v}(\rho)}$, $\alpha(\rho)$,$\rho_{A}(\rho)$ and $E_{kin}$.
Corrections due to the surface effects, which are necessary to estimate the associate bulk values have been evaluated using  the
following relation:
\begin{eqnarray}
  Q_{i}&=& Q_{b}+Q_{s}A_{i}^{-1/3}+0.45Q_{s}A_{i}^{-2/3}
\end{eqnarray}
$Q_{i}$ indicates the quantity valuated for the system with mass $A_{i}$ (i=1,2).
 $Q_{b}$ and $Q_{s}$ are the bulk and surface coefficients.
Effects related to the curvature are represented by the last term of eq.(42). The coefficient 0.45
has been deduced performing a couple of calculations in boxes having the same volume as the considered spheres.
  \begin{figure}
\includegraphics[scale=0.4]{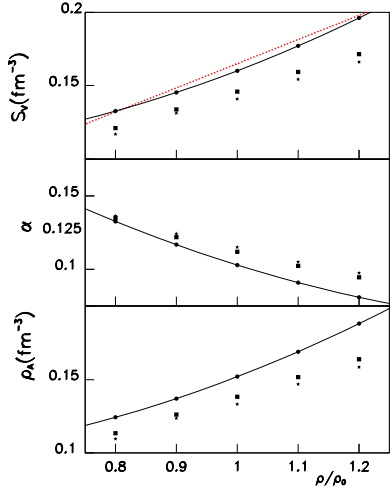}
\centering
\caption{\label{fig2} Typical result for the primary quantities $\overline{S_{v}(\rho)}$, $\alpha(\rho)$,$\rho_{A}(\rho)$ as a function of reduced density.
 The star symbols indicate the results of the study performed on the lighter system with mass $A_{1}$, the
 squares represent the results obtained for heavier system containing $A_{2}$ particles. Finally the circles
 indicate the obtained corrected values for surface effect according to eq.(42). The black solid lines are
 the final results of a fit procedure with a second order polynomial of the density . The red line
  represents the function $\rho$ as a function of the reduced density.}
\end{figure}

As an example, for $\beta=0$ and $\gamma=1$ (form factor $F_{2}$) in Fig.2 we show as a function
of the reduced density the values of $\overline{S_{v}(\rho)}$, $\alpha(\rho)$, $\rho_{A}(\rho)$
evaluated for the systems 1 (marked points with star), for the system 2
(marked points with square) and the bulk estimated values (dot points).
Corrections of the same order are evaluated also for the kinetic energy contribution.
In the following for simplicity we will refer to the bulk quantities without
using the subscript $b$.
After this first step, we fit the evaluated bulk quantities
with a polynomial function of the reduced density $\frac{\rho}{\rho_{0}}$.
The above quantities show in fact deviations from a
 simple linear behavior  as a function of $\rho$. This can be seen by looking at the red line in the figure ($\rho$ as  a function of $\rho/\rho_{0}$).
A second order polynomial reproduces very well the behavior in the range of explored densities.
The results of the fit are shown in the figures with lines.

The obtained functions are then substituted in $E^{C}$ and therefore the total binding energy can be now evaluated with continuity as a function of $\rho$ and  we can search for the parameter values  solving the CoMD functional $E^{C}$ obtained in Sec. II.1 and satisfying the conditions expressed
in eqs.(38-41).
We have to note that the numerical solution of this system of coupled equations in general can not be obtained with
the same precision as the one involving the functional $E$. In most of the cases, depending on the stiffness parameter $\gamma$ it is possible to obtain solutions reproducing the $E(\rho_{0})$ and $\rho_{0}$ values within $10\%$
while a larger spread is obtained for the $K_{NM}(\rho_{0})$. The chosen solutions will be
the one which minimize the total relative difference from the reference values.
\noindent
Having found the best solution for the functional $E^{C}$, in the sense above specified,
with the new set of parameter values for $T_{0}$, $T_{3}$ and
$\sigma$
 we perform another series of microscopic NM simulations on the systems of $A_{1}$,$A_{2}$ particles.
  After having included the correction for surface effects, we do the polynomial fit. Then using the new calculated quantities  we solve another time the functional
$E^{C}$.
This iterative procedure is continued until the values of the
parameters differs, in two subsequent  steps,  by an amount less than $\pm$5 \%.
The method  converges after 2-3 iterations.

\section{Results and discussion}
In this section we illustrate the results obtained from the recursive procedure previously described.
\subsection{Results on the primary quantities}
As an example in Fig.3 we show the final values of $\overline{S_{v}}$ as a function of the reduced density obtained in the case of symmetric NM and for different form factors $F_{k}$ ($T_{4}=32$ MeV). The lines represent the results of the fit with a second order polynomial. The red line represents instead the linear relation corresponding to the density $\rho$.
As we can see $\overline{S_{v}}$ shows deviations from $\rho$ depending also on the used form factor.
\begin{figure}
\includegraphics[scale=0.55]{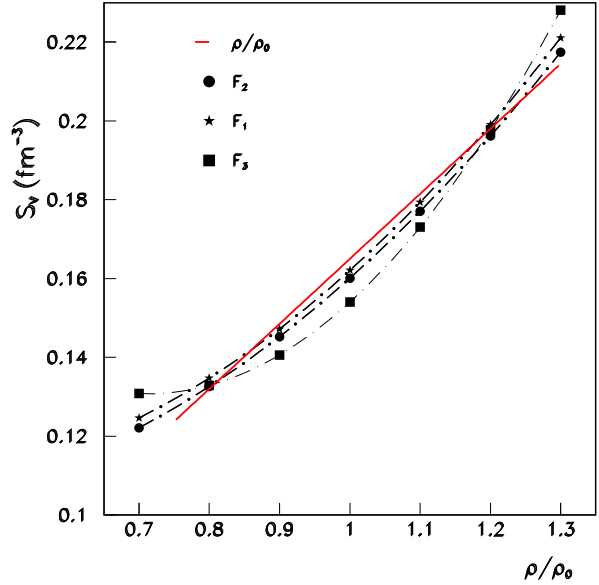}
\centering
\caption{\label{fig2} Final values of  $\overline{S_{v}(\rho)}$ for symmetric NM as a function of reduced density and for different form factors $F_{k}$ as indicated in the legend. The black solid lines are
 the result of a fit procedure with a second order polynomial of the density. The red line represent the density
 $\rho$}
\end{figure}

 Under the same conditions in Fig.4 with black line and points, we show the value of $\alpha$ as a function of the reduced density. The red line (and points) represents
the values of $\alpha$ obtained in the case of $T_{4}=0$ MeV  using the form factor $F_{2}$. The
blue line represents the  values  obtained for $T_{4}=59$ MeV which corresponds in the case of the
Se-MFA to a value of $e_{sym}(\rho_{0})$ equal to 42 MeV.
\begin{figure}
\includegraphics[scale=0.55]{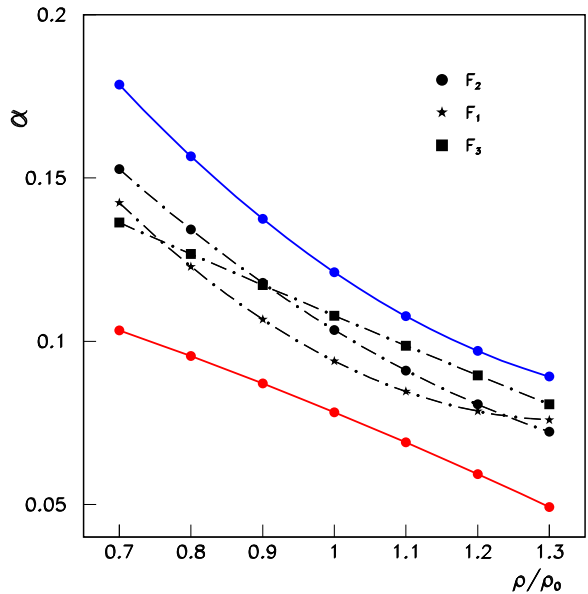}
\centering
\caption{\label{fig6} Final values of  the $\alpha$ correlation coefficient for $\beta=0$
as a function of reduced density. The black points and lines (results of a second order polynomial fit) refers to different form factors $F_{k}$ and $T_{4}=32$ MeV. For the forma factor $F_{2}$ the red and blu points refers to $T_{4}$ values
 equal to 0 and 59 MeV respectively.}
\end{figure}

As can be seen the value of $\alpha$ decreases with the density and increases with  $T_{4}$. For $T_{4}=0$
finite values of $\alpha$ are essentially due to correlations imposed by the Pauli principle in the system of interacting particles.

\subsection{Total binding energy}
In Fig.5, upper panel, we show with a solid line the total energy per nucleon $E$ as a function of the reduced
density  (eqs.(11,12)) satisfying the requested conditions on NM saturation properties.
This result represent our reference point.
\begin{figure}
\includegraphics[scale=0.55]{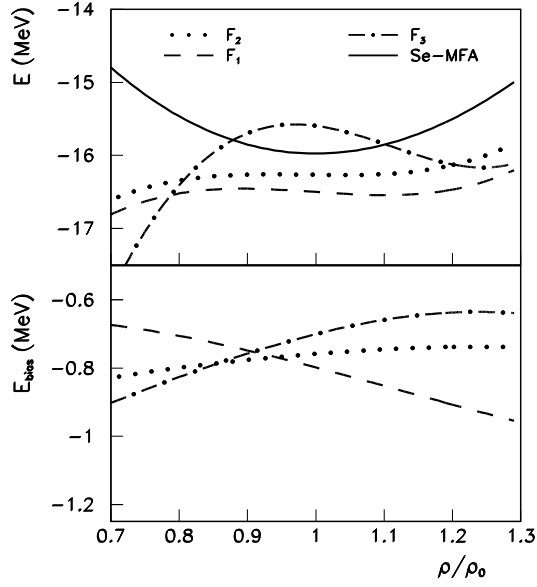}
\centering
\caption{\label{fig7} Upper panel:total energy per nucleon $E$ for symmetric NM as a function of the reduced density.The solid line represents the results obtained in the Se-MFA. With discontinue lines we plot the results for CoMD calculations corresponding to the first step of the iterative procedure (see the text). Different discontinue lines represents results related to different form factors $F_{k}$ according to the legend. Bottom panel: For the same parameter values the values
of $E_{bias}$ (see eq.(32)) are plotted as function of the reduced density.}
\end{figure}

The curves with broken lines represent  the results
of our NM simulations with CoMD model at the first step of the iteration for the different indicated form factors $F_{k}$. In this case the parameter used for
the effective interaction are the same like the ones obtained from the minimization of the energy functional
related to the Se-MFA. As can be seen the curves in this cases are rather different.
In particular for $F_{1}$ and $F_{2}$ the minimum energies are shifted to lower values and the compressibility modulus  also shows large deviations compared to the chosen reference value. $F_{3}$ shows instead even a maximum at the saturation density.
In the lower panel we show the corresponding values of $E_{bias}$ (see eq.(32)).
Together with the density
dependence  of the primary quantities above described, this term is the main co-responsible of the observed deviations.
The figure shows that the density behavior of  $E_{bias}$ strongly depends at a quantitative and qualitative level on the used form factors $F_{k}$. In particular we note an increasing slope from positive to negative values with the increasing of the degree of stiffness $\gamma$
 characterizing the behavior of the iso-vectorial forces.
We can expect therefore that the necessary correction  on the parameters
of the Iso-scalar effective interaction to reproduce the reference properties of the symmetric NM
will show a  dependence on $\gamma$.

In Fig.6 we show analogues results after that the self consistent iteration procedure
has been completed as described in Sec. III .
\begin{figure}
\includegraphics[scale=0.5]{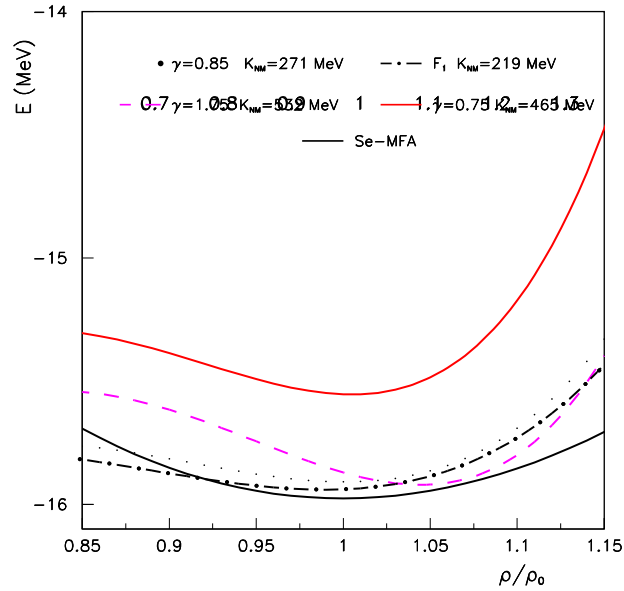}
\centering
\caption{\label{fig8} Total Energy per nucleon $E^{C}$ obtained through CoMD calculations for symmetric NM as a function of the reduced density. Different curves refer to different form factors $F_{k}$. In the legend the
value of $K_{NM}$ at $\rho_{0}$ are also indicated. The red and magenta lines represent
an example of limiting cases reached when $\gamma$ is well outside the range 1.5-0.85}
\end{figure}
 In the  interval  $\gamma \simeq 0.85-1.5$  the
saturation density and binding energies are reproduced within some percent. The compressibility is instead obtained within about $20\%$.
In the figure we show some example of these solutions with black lines.
In general we note a more pronounced asymmetry of the curves around the saturation density with a reduced slope
of the lower density branch and an increased slope for reduced density larger than 1.
This behaviors is a direct consequence of the density dependence of the average overlap integral (see Fig.5).
For stiffness parameter values out of the indicated interval we observe a fast increasing of the compressibility
(well beyond 400 MeV) and of the corresponding binding energy at the saturation density.
In Fig.6 an example showing this trend is represented by colored lines obtained
for $\gamma =0.75$ and $\gamma=1.75$.
 All these circumstances shows that, in  the frame work of the present molecular dynamics approach,
 the $F_{4}$ form factor with stiffness parameter values external to  the range
  $\gamma\approx 1.5-0.85$, due to the correlations discussed, is not able
  to reproduce the chosen commonly accepted saturation properties of the symmetric nuclear matter.
These results are consistent with recent findings on the study
 of the $^{40,48}Ca+^{40,48}Ca$ systems at 25 MeV/nucleon \cite{lim,lim1} concerning the balance
 between the yields of incomplete-fusion
 and multi-breakup processes.
Finally in Fig.7 we show as a function of the $\gamma$ parameter the values of the $T_{0}$, $T_{3}$ and $\sigma$  obtained from the iterative procedure described in
Sec. III.1.
Apart from the extremal plotted values, corresponding to high values of the compressibility ( see Fig.6), the
 internal ones correspond to saturation density, binding energies and $K_{NM}$ values well within 20\% of the value obtained in the case of the Se-MFA.
 From the figure we observe a dependence of the  parameter values describing the iso-scalar forces on the stiffness parameter $\gamma$ associated to the iso-vectorial interaction. In the internal region
 of the explored interval, even if the dependence is moderate, maximum changes of the order of 16$\%$ and $20\%$
 are obtained for the $T_{3}$ and $\sigma$ values respectively.
 However it is remarkable that the average values of the iso-scalar interaction parameters show large differences
 compared to the reference values obtained in the case of the Se-MFA ($T_{0}=-263$ MeV,$T_{3}=208$ MeV and
 $\sigma=1.25$. see Sec.III) which are instead independent on $\gamma$.
 \begin{figure}
\includegraphics[scale=0.5]{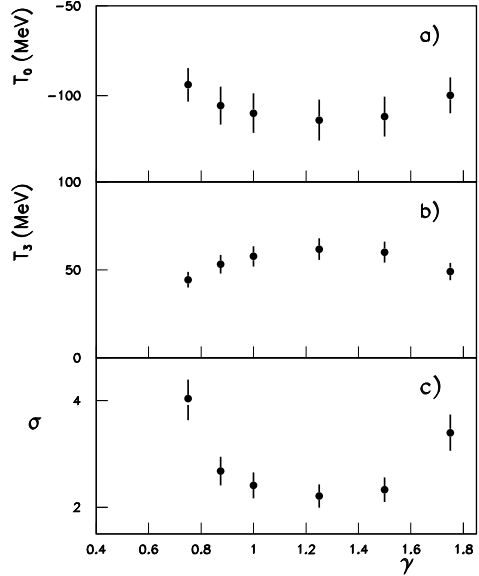}
\centering
\caption{\label{fig9} The values of the parameters $T_{0}$ $T_{3}$ and $\sigma$ (panel a,b,c respectively)
obtained through the iterative procedure  applied to CoMD calculations (see the Sec. III.1)are shown as a function of the stiffness parameter $\gamma$. The values of $\gamma$ equal to 1 and 1.5 are associated to the functional form $F_{2}$ and $F_{1}$ respectively (see Sec.II). The others ones are instead associated to $F_{4}$. The bar errors represent the global uncertainty on the parameter values related to the numerical procedures}
\end{figure}

\section{Summary and Concluding Remarks}
Many-body correlations produced in Molecular Dynamics approach based on the CoMD model have been discussed
and their connection with the used effective interaction have been
analyzed in the case of symmetric nuclear matter simulations.
This study has been performed by comparing the results obtained for the total energy functional and
the nuclear matter (NM) main saturation properties with
the ones obtainable in the case of a semiclassical mean field approximations (Se-MFA). This comparison has been
performed by  using the same kind of simple effective Skyrme interaction.
While we can expect the two approaches to produce large differences at low
density due to the cluster formation process, the following study shows that noticeable differences
are obtained also in a narrow range of densities around the saturation one where no cluster production
has been observed.
The effective Skyrme interaction include two-body and three-body effective iso-scalar interactions plus a
two-body iso-vectorial one with form factors commonly used also in Se-MFA.
In  CoMD model calculations, around the saturation density, the effects related to the spacial correlations generated by the usage of the localized wave-packets and the ones associated to  the multi-particles
correlations produced through the Pauli principle constraint can be well described through a second order
polynomial decomposition of the total energy as a function of the density.
 The obtained results show that, contrary to the case of the Se-MFA,
the discussed correlations produce an interdependence between the parameters describing the iso-scalar forces and
the ones related to the iso-vectorial interaction.
The usage of an iterative procedure tuned to obtain in both the cases (Se-MFA and in CoMD approaches) very similar saturation density, total
energy and compressibility values for symmetric NM allowed to extract the "good" set of parameters  used for CoMD calculations.
The  obtained values differ under many aspects from the ones obtained from the Se-MFA:
in particular the density dependence of the used form factors describing the iso-vectorial
forces has to change now in a more restricted range of stiffness values.
Moreover the values of the
coefficients describing the iso-scalar interactions are rather different in the two cases.
Work is in progress to extend these studies to asymmetric NM.
Finally we conclude by observing that even if from a numerical point of view the obtained results are strictly
valid for the CoMD model, the performed study  shows that the observed differences in the
parameter values describing the chosen effective interaction can have a more wide meaning. They are indeed strictly linked to some general properties of the semiclassical wave packets dynamics.

\ack
We wish to acknowledge  discussions with U.Lombardo, G.Giuliani, A.Bonasera and
V.Greco

\end{document}